\documentstyle[aps]{revtex} 

\input epsf         
\epsfverbosetrue    

\begin{document}


\draft

\title{Universal zero-frequency Raman slope in a $d$-wave superconductor}

\author{W. C. Wu and J. P. Carbotte}
\address{Department of Physics and Astronomy, McMaster University\\
Hamilton, Ontario, Canada L8S 4M1}
 
\date{\today}

\maketitle

\begin{abstract}
It is known that for an unconventional superconductor with nodes in the 
gap, the in-plane microwave or dc conductivity saturates at low temperatures
to a universal value independent of the impurity concentration.
We demonstrate that a similar feature can be accessed using 
channel-dependent Raman scattering. 
It is found that, for a $d_{x^2-y^2}$-wave superconductor,
the slope of low-temperature Raman intensity at zero frequency is 
universal in the $A_{1g}$ and $B_{2g}$ channels, but not in the 
$B_{1g}$ channel.  Moreover, as opposed to the microwave 
conductivity, universal Raman slopes are sensitive not only
to the existence of a node, but also to different pairing 
states and should allow one to distinguish
between such pairing states.
\end{abstract}

\vskip 0.2 true in


The effect of impurity scattering has been
of increasing interests in studies of High-$T_c$ 
superconductors \cite{Lee93,HPS,AMC93,HG93,SM95,GYSR96}.
For superconductors with an order parameter which has nodes
on the Fermi surface, it is well known that impurity scattering can 
lead to a finite density of quasiparticle states at zero energy
(gapless excitations). In particular, in the strong resonant scattering 
limit, quasiparticle states can be strongly localized as a result of 
the short coherence length and mean free path \cite{Lee93}
provided that the impurity concentration is low.
This has some interesting observable consequences.
Of equal interest is that, as first predicted by Lee \cite{Lee93},
while the effective impurity scattering rate $\gamma$ is quite different
for different impurity concentrations and different scattering limits
(for example, $\gamma\sim \Delta_0 e^{-\Delta_0 \tau}$ in the Born limit and
$\gamma\sim \Delta_0 (\Delta_0 \tau)^{-{1\over 2}}$ in the unitary limit with
$1/\tau$ the normal-state scattering rate and $\Delta_0$ the maximum of the
gap), the microwave conductivity 
saturates at low temperatures ($\sigma_0 =ne^2/\pi m\Delta_0$)
and is independent of $\gamma$ (or the impurity concentration).
The experimental verifications of this universal feature
gives unambiguous evidence that the
order parameter in High-$T_c$ materials exhibits nodes on the Fermi surface.

In the context of thermal conductivity,
Graf {\em et al.} \cite{GYSR96} have found a similar
universality related to the microwave conductivity 
via the Wiedemann-Franz law. 
This has been confirmed in an experiment by Taillefer {\em et al.}
\cite{TLGBA97} which measures the in-plane low-temperature thermal 
conductivity of YBa$_2$Cu$_3$O$_{6.9}$ at different Zn substitutions for Cu.
In the present letter, we demonstrate how one can also study these universal 
features by doing channel-dependent Raman scattering experiments. It is 
found, in a $d_{x^2-y^2}$-wave superconductor, 
that the low-temperature slope of
the Raman intensity at zero frequency are universal in the
$A_{1g}$ and $B_{2g}$ channels, but strongly dependent 
on the scattering rate ($\sim \gamma^2$) in the $B_{1g}$ channel. 
As opposed to the microwave or
thermal conductivity for which an entire Fermi surface average is
taken and thus the universal feature is general for all the pairing states
so long as gap nodes cross the Fermi surface
(except for a scale factor), 
which channels saturate and are universal and which do not
saturate in Raman scattering change with the gap symmetry.
This channel-dependent universality exhibited in the Raman 
scattering thus gives one more method to test the symmetry of the
pairing state in the High-$T_c$ superconductors and
is more powerful than microwave or thermal conductivity
because of the additional selectivity involved in the Raman geometry
which allows some information on the
position of the nodes in the Brillouin zone to be obtained.
This phenomenon should be of particular interest
in the heavy fermion superconductors 
where the pairing states are considered to be more diverse.

The saturation of the low-temperature
microwave conductivity in a $d$-wave superconductor, is 
a result of a cancellation between the value of the impurity-induced 
density of states at zero energy and the 
quasiparticle relaxation lifetime and arises only if
there exists nodes in the order parameter on the Fermi surface.
The saturation in the slope of the
low-temperature zero-frequency Raman intensity 
can be understood in a similar manner, with an additional
channel dependent feature unique to Raman which arises from the different 
dependence on the chosen Raman geometry
vertices which pick up different contributions around the Fermi surface.
In a $d_{x^2-y^2}$-wave superconductor,
the $B_{2g}$-channel Raman vertex selects preferentially
states along the diagonals of the Brillouin zone
(where the nodal lines are) and hence
probes directly the low-lying quasiparticle excitations. Consequently
the net result is 
similar to what is seen in the microwave conductivity case.
In contrast in the $B_{1g}$ channel, the Raman vertex has maximum weight 
along the $k_x$ or $k_y$ axes and zero weight along the diagonals,
consequently one is effectively measuring a finite gap and no
universality is observed.

The Raman intensity is proportional to the imaginary part of 
the zero-momentum limit generalized density response function 
 
\begin{eqnarray}
\chi_{\gamma\Gamma}(i\nu_n)=
-T\sum_{{\bf k},\omega_{n}}
{\rm Tr} [\hat{\gamma}({\bf k})
\hat{G}({\bf k},i\omega_{n}+i\nu_n))\hat{\Gamma}({\bf k},i\nu_n)
\hat{G}({\bf k},i\omega_{n})],
\label{eq:chi.Raman}
\end{eqnarray}
where Tr denotes a trace and $\hat{\gamma}({\bf k})=
\hat{\tau}_3\gamma({\bf k})$ is the bare Raman vertex with
$\hat{\tau}_3$ the Pauli matrix and
$\gamma({\bf k})={m\over \hbar^{2}}{\bf e}_s \cdot {\partial^2\xi_{\bf k}
\over \partial k_s\partial k_i} \cdot{\bf e}_i$ 
(effective mass approximation \cite{AG74}).
Here $\xi_{\bf k}$ is the electronic dispersion relation of 
the superconducting layer and ${\bf e}_i$ (${\bf e}_s$)
correspond to the polarizations of incident (scattered) photons.
The renormalized Raman vertex $\hat{\Gamma}$
in (\ref{eq:chi.Raman}) is given by

\begin{eqnarray}
\hat{\Gamma}({\bf k},i\nu_n)=\hat{\tau}_3\gamma({\bf k})-\hat{\tau}_3 
v_c T\sum_{{\bf k}^\prime,\omega_{n}} {\rm Tr} [\hat{\tau}_3
\hat{G}({\bf k}^\prime,i\omega_{n}+i\nu_n)\hat{\Gamma}({\bf k}^\prime,i\nu_n)
\hat{G}({\bf k}^\prime,i\omega_{n})],
\label{eq:Gamma.Coulomb}
\end{eqnarray}
where $v_c$ is the Coulomb interaction.
In Eq.~(\ref{eq:Gamma.Coulomb}), we have ignored the contribution
to the vertex corrections
due to the impurity potentials and two-particle pairing interactions
and have included only the effect of the Coulomb interaction.
For isotropic impurity scattering,
it is sufficient to use the bubble diagram at small ${\bf q}$, 
while the inclusion of the pairing interaction vertex correction is 
shown to have little effect on the Raman spectra \cite{Dev95} and is
particularly negligible at the low frequencies of interest.
However, the effect of impurities is fully included in the
single-particle Green's function $\hat{G}$ in Eqs.~(\ref{eq:chi.Raman}) 
and (\ref{eq:Gamma.Coulomb}).
Substituting Eq.~(\ref{eq:Gamma.Coulomb}) into (\ref{eq:chi.Raman}),
one obtains

\begin{eqnarray}
\chi_{\gamma\Gamma}(i\nu_n)&=&
\chi_{\gamma\gamma}(i\nu_n)-{\chi_{\gamma 1}(i\nu_n)
\chi_{1\gamma}(i\nu_n)\over
\chi_{11}(i\nu_n)-v_c^{-1}},
\label{eq:chi.Raman.screened}
\end{eqnarray}
where $\chi_{\gamma 1}(i\nu_n)$ is defined in the same way as
$\chi_{\gamma\Gamma}(i\nu_n)$ with $\hat{\Gamma}$ 
replaced by $\hat{\tau}_3$ in (\ref{eq:chi.Raman}), and so on.

In terms of the particle-hole space, the single-particle Green's function 
is given by
$\hat{G}^{-1}({\bf k},i\omega_n)= i\tilde{\omega}_n\hat{\tau}_0-
\tilde{\xi}_{\bf k}\hat{\tau}_3-\tilde{\Delta}_{\bf k}\hat{\tau}_1$, where
$\tilde{\omega}_n$, $\tilde{\xi}_{\bf k}$, and $\tilde{\Delta}_{\bf k}$
are the impurity-renormalized Matsubara frequencies, 
electron energy spectrum,
and gap. $\hat{G}$ is related to the noninteracting 
Green's function $\hat{G}_0^{-1}({\bf k},i\omega_n)= i{\omega}_n\hat{\tau}_0-
{\xi}_{\bf k}\hat{\tau}_3-{\Delta}_{\bf k}\hat{\tau}_1$
via the Dyson's equation $\hat{G}^{-1}({\bf k},i\omega_n)
=\hat{G}_0^{-1}({\bf k},i\omega_n)-\hat{\Sigma}({\bf k},i\omega_n)$.
We shall solve the self-energy 
$\hat{\Sigma}$ due to the impurity scattering.
By expanding $\displaystyle \hat{\Sigma}(i\omega_n)\equiv \sum_{\alpha}
\Sigma_\alpha (i\omega_n) \hat{\tau}_\alpha$ ($\alpha=0,1,3$),
one finds $i\tilde{\omega}_n=i\omega_n-\Sigma_0$,
$\tilde{\xi}_{\bf k}={\xi}_{\bf k}+\Sigma_3$,
and  $\tilde{\Delta}_{\bf k}=\Delta_{\bf k}+\Sigma_1$.
Employing the usual $T$-matrix approximation, the self-energy is then
given by $\hat{\Sigma}({\bf k},i\omega_n)=n_i 
\hat{T}({\bf k,k},i\omega_n)$, where
$n_i$ is the impurity density and
 
\begin{eqnarray}
\hat{T}({\bf k,k^\prime},i\omega_n)=
v_i({\bf k,k^\prime})\hat{\tau}_3+\sum_{\bf k^{\prime\prime}}
v_i({\bf k,k^{\prime\prime}})\hat{\tau}_3\hat{G}({\bf k^{\prime\prime}},
i\omega_n)
\hat{T}({\bf k^{\prime\prime},k^\prime},i\omega_n).
\label{eq:t.matrix}
\end{eqnarray}
Here $v_i({\bf k,k^\prime})\equiv
\langle{\bf k^\prime}|v_i|{\bf k}\rangle$ is the impurity potential.
If we consider only isotropic impurity scattering 
[$v_i({\bf k,k^\prime})= v_i$], the $T$-matrix in (\ref{eq:t.matrix}) is 
left only with frequency dependence and can be solved to get
$\hat{T}(i\omega_n)= [1-v_i\hat{\tau}_3
\hat{\overline{G}}(i\omega_n)]^{-1} v_i\hat{\tau}_3$
with the integrated Green's function $\hat{\overline{G}}(i\omega_n)\equiv
\sum_{\bf k}\hat{G}({\bf k}, i\omega_n)$.
One can expand $\displaystyle
\hat{\overline{G}}(i\omega_n)=\sum_{\alpha} G_\alpha(i\omega_n)
\hat{\tau}_\alpha$ ($\alpha=0,1,3$) with
$G_\alpha(i\omega_n)\equiv 1/2 \sum_{\bf k}{\rm Tr}[\hat{\tau}_\alpha
\hat{G}({\bf k},i\omega_n)]$.
For a superconductor with particle-hole symmetry
and an odd-parity gap which is the case for the $d_{x^2-y^2}$ state,
$G_1(i\omega_n)=G_3(i\omega_n)=0$.  This immediately gives 
 
\begin{eqnarray}
\Sigma_0={n_i G_0\over c^2- G_0^2}, ~~~~~~~
\Sigma_1=0, ~~~~~~~
\Sigma_3={c n_i \over c^2- G_0^2},
\label{eq:Sigmas.1}
\end{eqnarray}
where $c\equiv 1/v_i$.
The result of $\Sigma_1=0$ is a reflection of the well-known result
that in an unconventional superconductor with nodes in the gap
and zero average, the gap remains unrenormalized
due to the impurity scattering ($\tilde{\Delta}_{\bf k}=\Delta_{\bf k}$).
Furthermore the effect of $\Sigma_3$ is absorbed
into the chemical potential as usual and consequently 
$\tilde{\xi}_{\bf k}\equiv \xi_{\bf k}$.
Equation~(\ref{eq:Sigmas.1}) is convenient for discussing
both the weak (Born) scattering ($c\gg 1$) 
and strong (resonant) scattering ($c\ll 1$) limit.
In the normal state (${\Delta}_{\bf k}=0$), one can easily work out that
$i\tilde{\omega}_n=i\omega_n+i(1/2\tau)$, where
in the Born limit,  the (isotropic) scattering rate 
$1/2\tau=2\pi N(0)n_i v_i^2$, while in the
resonant limit,  $1/2\tau=n_i/2\pi N(0)$. Here
$N(0)=m/2\pi\hbar^2$ is the density of states per spin
on the Fermi surface.

One can use a general Raman vertex
$\gamma_s({\bf k})=\gamma_s^0+\gamma_s^1 f_s(\phi)$ to classify different
symmetry channels \cite{Dev95,Dev94-3} denoted by $s$.
Here $\gamma_s^0$ represents the
isotropic and $\gamma_s^1$ represents the anisotropic part of
$\gamma_s$ and $\phi$ is the azimuthal angle in the $x$-$y$ plane. 
In the case of a cylindrical Fermi surface, 
$f_s(\phi)=\cos(2\phi)$ for the $B_{1g}$ channel,
$f_s(\phi)=\sin(2\phi)$ for the $B_{2g}$ channel, and 
$f_s(\phi)=\cos(4\phi)$ for the $A_{1g}$ channel. 
With the above Raman vertices and  
in the case of perfect screening ($v_c^{-1}\rightarrow 0$),
Eq~(\ref{eq:chi.Raman.screened}) is reduced to

\begin{eqnarray}
\chi_{\gamma\Gamma}(i\nu_n)=(\gamma_s^1)^2
\left[\chi_s^{2}(i\nu_n)-{[\chi_s^{1}(i\nu_n)]^2
\over \chi_s^{0}(i\nu_n)}\right],
\label{eq:chi.Raman.sc.simplified}
\end{eqnarray}
where we have defined

\begin{eqnarray}
\chi_s^{i}(i\nu_n)=
-T\sum_{{\bf k},\omega_{n}}
[f_s(\phi)]^i{\rm Tr} [\hat{\tau}_3
\hat{G}({\bf k},i\omega_{n}+i\nu_n))\hat{\tau}_3
\hat{G}({\bf k},i\omega_{n})].
\label{eq:chi.i}
\end{eqnarray}
In (\ref{eq:chi.Raman.sc.simplified}), the isotropic term ($\gamma_s^0$) 
is dropped as it cancels due to Coulomb screening.
In the following we shall limit ourselves to a 
$d_{x^2-y^2}$-wave superconductor with gap 
$\Delta_{\bf k}=\Delta_0\cos(2\phi)$.  In both the cases of the
$B_{1g}$ and $B_{2g}$ channels, the second term in
(\ref{eq:chi.Raman.sc.simplified}) vanishes since
$\chi_s^{1}(i\nu_n)=0$. Technically this is due to
the Fermi surface average $\langle f_s(\phi) |\Delta_{\bf k}|^2\rangle=0$.
However this condition doesn't hold in the $A_{1g}$ channel
where the squared gap function has a component identical to
$f_{A_{1g}}(\phi)=4\cos(4\phi)$ and hence
$\langle \cos(4\phi) |\Delta_{\bf k}|^2\rangle\neq 0$.
As a consequence, the Coulomb screening only has an effect on
the $A_{1g}$ channel intensity and has no effect on the $B_{1g}$
and $B_{2g}$ channels. 

Based on the cylindrical Fermi surface approach,
the second term of Eq.~(\ref{eq:chi.Raman.sc.simplified}) 
in the $A_{1g}$ channel is shown to be of the same order
as the first term in the zero-frequency limit \cite{Wu97}.
In general, however, the effect of the second 
term is quite sensitive to the underlying quasiparticle energy
dispersion relation and the issue regarding the effect of screening
in the $A_{1g}$ channel remains an issue of considerable
debate \cite{KC94}. 
On the other hand, experimental data seems to suggest
that the first term of Eq.~(\ref{eq:chi.Raman.sc.simplified})
can account well for the Raman intensity in the low-frequency regime.
We thus drop the second term
of (\ref{eq:chi.Raman.sc.simplified}) in our calculations
of the slopes of the {\em channel dependent} Raman intensity
at zero frequency in {\em all} channels. These are defined
by ($i\nu_n\rightarrow \Omega+i\delta$)

\begin{eqnarray}
S\equiv \left.{d\chi^{\prime\prime}_{\gamma\Gamma}(\Omega)
\over d\Omega}
\right|_{\Omega\rightarrow 0}=
\left.{\chi^{\prime\prime}_{\gamma\Gamma}(\Omega)\over \Omega}
\right|_{\Omega\rightarrow 0},
\label{eq:def.S}
\end{eqnarray}
where the double prime denotes an imaginary part and
the equal sign arises because
$\chi^{\prime\prime}_{\gamma\Gamma}(\Omega=0) =0$.
Using the spectral representation for the imaginary frequency
Green's function, analytically continuing to real frequency
from imaginary frequency ($i\omega_n\rightarrow \omega+i\delta$),
and then performing the frequency sum and 
momentum sum (replaced by an integration $\sum_{\bf k}=
2N(0)\int_{-\infty}^\infty d\xi\int_0^{2\pi} {d\phi\over 2\pi}$) gives

\begin{eqnarray}
S&=&
N(0)[\gamma_s]^2\int_0^{2\pi} {d\phi\over 2\pi}[f_s(\phi)]^2
\int_{-\infty}^\infty d\omega{f(\omega)-f(\omega-\Omega)\over \Omega}
\raisebox{0.8cm}{~~~}\nonumber\\
&\times& {\rm Im}\left[
{\tilde{\omega}_+^\prime(\tilde{\omega}_+
+\tilde{\omega}_+^\prime)-2\Delta_{\bf k}^2\over
({\xi}_+ +{\xi}_+^\prime){\xi}_+
{\xi}_+^\prime}
+
{\tilde{\omega}_-^\prime(\tilde{\omega}_+
+\tilde{\omega}_-^\prime)-2\Delta_{\bf k}^2\over
({\xi}_+ -{\xi}_-^\prime){\xi}_+
{\xi}_-^\prime}
\right]_{\Omega\rightarrow 0}\raisebox{1.0cm}{~~~}
\label{eq:chi.reduced}
\end{eqnarray}
where $\tilde{\omega}_\pm\equiv i\tilde{\omega}_n(\omega\pm i\delta)$;
$\tilde{\omega}_\pm^\prime\equiv i\tilde{\omega}_n(\omega-\Omega\pm i\delta)$
and ${\xi}_\pm\equiv {\rm sgn}(\omega)\sqrt{\tilde{\omega}_\pm^2
-\Delta_{\bf k}^2}$; ${\xi}_\pm^\prime\equiv {\rm sgn}(\omega-\Omega)
\sqrt{(\tilde{\omega}_\pm^\prime)^2-
\Delta_{\bf k}^2}$ which are chosen to have branch cuts such that
${\rm Im}\xi_+, {\rm Im}\xi_+^\prime>0$ and
${\rm Im}\xi_-, {\rm Im}\xi_-^\prime<0$. 
The index $1$ in the vertex is dropped 
($\gamma_s^1\rightarrow \gamma_s$) for simplicity.

It is useful to compare Eq.~(\ref{eq:chi.reduced}) for the zero-frequency
Raman slope with a similar expression for the microwave conductivity
(see, for example, Eq.~(2) of Ref.~\cite{HPS}).
One finds that they are the same apart from an overall constant factor
which appears in front of the expression and from a different
angular function [in Raman scattering, the angular function
$f_s(\phi)=\cos(4\phi)$, $\cos(2\phi)$, or $\sin(4\phi)$ for
$A_{1g}$, $B_{1g}$, and $B_{2g}$ channels,
while in the conductivity the angular function is usually 
$\hat{p}_x=\cos(\phi)$ for calculating $\sigma_0^{xx}$ or 
$\hat{p}_y=\sin(\phi)$ for calculating $\sigma_0^{yy}$]. Also the 
appearance of
term $\Delta_{\bf k}^2$ in the second line of (\ref{eq:chi.reduced})
is unique to Raman and occurs due to the
different type of vertex (which is coupled to $\hat{\tau}_3$ in Raman
and $\hat{\tau}_0$ in the conductivity).
The term proportional to $\Delta_{\bf k}^2$, however,
will drop out due to a cancellation between the two terms
at zero temperature and contributes only a small amount
at finite temperatures.

We consider first the $T=0$ limit which gives
$[f(\omega)-f(\omega-\Omega)]/\Omega\approx \partial f(\omega)/
\partial \omega\approx -\delta(\omega)$ when $\Omega\rightarrow 0$.
This means that the $\omega$ integration in (\ref{eq:chi.reduced})
is sharply peaked around the small $\omega$ region centered at
$\omega=0$. Consequently we have

\begin{eqnarray}
S=N(0)\gamma_s^2\left\langle
{[f_s(\phi)]^2\gamma^2\over  
(\gamma^2+\Delta_{\bf k}^2)^{3\over 2}}
\right\rangle,
\label{eq:S.avg}
\end{eqnarray}
where $\langle\cdot\cdot\cdot\rangle$ denotes an 
average over the Fermi surface. This expression agrees with one
obtained before by Devereaux and Kampf \cite{DK96}.
Here the impurity-induced
scattering rate in the superconducting state
at zero-frequency is $\gamma= -i\tilde{\omega}(\omega=0)=
i\Sigma_0(\omega=0)$.
The self-consistent results for $\gamma$ in the two different scattering
limits Born and resonant were solved for
by Lee \cite{Lee93} as mentioned earlier.
Assuming that $\gamma\ll \Delta_0$ (which requires that the
impurity concentration $n_i$ be small in the resonant limit), 
we find at $T=0$

\begin{eqnarray}
\left.
\begin{array}{rll}
S~ \sim~& \displaystyle {m \gamma_{s}^2 \over \pi^2 \hbar^2\Delta_0}
&\hspace{0.8cm}\makebox[3.5cm]{\rm $s=B_{2g}$ or $A_{1g}$}
\raisebox{0.8cm}{~~~}\\ 
\sim~& \displaystyle {m \gamma_{s}^2 \over \pi^2 \hbar^2\Delta_0}
\left({\gamma\over \Delta_0}\right)^2\ln\left({\Delta_0\over\gamma}\right)
&\hspace{0.8cm}\makebox[3.5cm]{\rm $s=B_{1g}$.}\raisebox{1.0cm}{~~~}\\ 
\end{array}
\right.
\label{eq:Raman.slopes}
\end{eqnarray}
As shown clearly in (\ref{eq:Raman.slopes}),
the  zero-frequency Raman slopes in both $B_{2g}$ and $A_{1g}$-channel 
exhibit a universal saturated value at $T=0$ which is independent of $\gamma$,
{\em i.e.}, of impurity concentration -- a feature first discovered by 
Lee \cite{Lee93} in the context of the microwave conductivity. 
The reason the $B_{2g}$ and $A_{1g}$ channels share the same limiting
value is simply because the square of angular functions
$[f_{B_{2g}}(\phi)]^2=\sin^2(2\phi)=1-\cos^2(2\phi)$
and $[f_{A_{1g}}(\phi)]^2=\cos^2(4\phi)=[1-2\cos^2(2\phi)]^2$
and the contribution due to whatever terms couple to $\cos(2\phi)$ is small.
In contrast in the $B_{1g}$ channel, the zero-frequency Raman slope 
is proportional to $\gamma^2$ (up to a logarithmic correction)
and hence is strongly dependent on the impurity concentration.
In the unitary limit, $\gamma\sim\tau^{-1/2}\sim n_i^{1/2}$,
therefore the $B_{1g}$ slope $S\sim n_i$.
We recall that for a system with tight-binding bands,
the Raman vertex strength $\gamma_{A_{1g}}$ and $\gamma_{B_{1g}}$ in 
(\ref{eq:Raman.slopes}) is proportional to the nearest-neighbor hopping, 
while $\gamma_{B_{2g}}$ is proportional to the next nearest-neighbor hopping
\cite{Dev94-1}.

If we take the ratio between the Raman intensities from
superconducting and normal states, we find in the limits
of $\Omega=0$ and $T=0$, 
$\chi_S^{\prime\prime}/\chi_N^{\prime\prime}
\sim \gamma^2/\Delta_0^3\tau$ in the $B_{1g}$ channel and
$\sim 1/\Delta_0\tau$ in the $B_{2g}$ channel. 
These differ from the expression given by 
Devereaux and Kampf \cite{DK96} in that they
use $\gamma$ for $1/\tau$ in the normal state.

The finite but low-temperature ($T\alt \gamma$) limit is obtained by
expanding $\tilde{\omega}_\pm=\pm i(\gamma+b\omega^2)+a\omega$
at small $\omega$ in (\ref{eq:chi.reduced}), where $\gamma$, $a$ and $b$
are constants and are found to be
$a\simeq 1/2$ and $b\simeq 1/(8\gamma)$ in the resonant scattering limit of 
primary interest here. Expanding the integrand in 
(\ref{eq:chi.reduced}) to second order in $\omega$ leads to the finite-$T$
result

\begin{eqnarray}
\left.
\begin{array}{rll}
S~ \sim~ & \displaystyle{m \gamma_{s}^2 \over \pi^2 \hbar^2\Delta_0}
\left(1+{\pi^2\over 36}{T^2\over \gamma^2}\right)
&\hspace{0.8cm}\makebox[3.5cm] {\rm $s=B_{2g}$ or $A_{1g}$}
\raisebox{0.8cm}{~~~}\\
\sim~ & \displaystyle{m \gamma_{s}^2 \over \pi^2 \hbar^2\Delta_0}
\left({\gamma\over \Delta_0}\right)^2\ln\left({\Delta_0\over\gamma}\right)
\left(1+{\pi^2\over 12}{T^2\over \gamma^2}\right)
&\hspace{0.8cm}\makebox[3.5cm] {\rm $s=B_{1g}$.}\raisebox{1.0cm}{~~~}\\
\end{array}
\right.
\label{eq:Raman.slopes.T}
\end{eqnarray}
While the universality is channel dependent, the variation
$\sim T^2$ is found in all three channels.  Equivalent results 
were given by Hirschfeld {\em et al.} \cite{HPS} 
for the microwave conductivity and by 
Graf {\em et al.} \cite{GYSR96} for the thermal conductivity.

As mentioned before, the channel-dependent universal zero-frequency
Raman slopes are sensitive to differences in pairing states.
For completeness, we consider some other pairing states
of interest and focus only on the $B_{1g}$ and $B_{2g}$ channels at $T=0$. 
The opposite result to (\ref{eq:Raman.slopes}) with $B_{1g}$ and $B_{2g}$ 
channels switched, is obtained if the gap has $d_{xy}$ symmetry 
with $\Delta_{\bf k}=\Delta_0\sin(2\phi)$.
In this case, one finds a universal feature in the
$B_{1g}$-channel Raman slope, but not in the $B_{2g}$ channel.
This is because one sees mainly regions of maximum gaps
in the $B_{2g}$-channel, while in the $B_{1g}$-channel, one sees mainly 
the regions of zero gap.
For a system with extended $s$-wave pairing state, {\rm i.e.},
with $\Delta_{\bf k}=\Delta_0\cos(4\phi)$, we find that
both $B_{1g}$ and $B_{2g}$ channels have the same universal
Raman slope $S=m\gamma_{s}^2 /2\pi^2 \hbar^2\Delta_0$.
In this case, the gap nodes appear at the angles of $\pi/8$, $3\pi/8$ and
the equivalent and consequently both $B_{1g}$ and $B_{2g}$ channels see
the same effective contribution from the gap node regions. 

Finally, we consider the case of a mixing gap with a $d_{x^2-y^2}$-wave
component and a {\em small} isotropic $s$-wave component 
of weight $\alpha$ with the gap given by
$\Delta_{\bf k}=\Delta_0[\cos(2\phi)+\alpha]$, where $\alpha<1$.
Physically this is somewhat similar to
a system with orthorhombic band structure and a pure $d_{x^2-y^2}$-wave
gap. In contrast to the pure $d_{x^2-y^2}$-wave case,
the gap nodes are now shifted away from the diagonals 
and consequently the $B_{1g}$ Raman channel will have some contribution 
from the node regions. This means that the universal zero-frequency
Raman slope feature will also be present in the $B_{1g}$ channel
and its weight will depend on how far the gap 
node have shifted off the diagonals, {\em i.e.}, how much
node contribution this channel picks up.  To leading order, we find
$S=m \gamma_{s}^2/ \pi^2 \hbar^2\Delta_0$
in the $B_{2g}$ channel (same as the pure $d_{x^2-y^2}$-wave case), while
in the $B_{1g}$ channel,

\begin{eqnarray}
S_{B_{1g}}= {m \gamma_{s}^2 \over \pi^2 \hbar^2\Delta_0}
\left[\alpha^2+ \gamma^2/\Delta_0^2\ln(\Delta_0/\gamma)\right].
\label{eq:Ramanslope.s+d}
\end{eqnarray}
Therefore when $\alpha\gg \gamma/\Delta_0$, the $B_{1g}$ Raman slope has
universality which however breaks down when $\alpha\alt \gamma/\Delta_0$.
In the unitary limit, $\gamma\sim n_i^{1/2}$ and thus
$S_{B_{1g}}\sim m\gamma_{s}^2 /\pi^2 \hbar^2\Delta_0 (\alpha^2+c n_i)$. 
A study of the impurity concentration dependence of $S_{B_{1g}}$
may be used to extract the size of $\alpha^2$ which in turn
enables one to test how orthorhombic the system is.  
We note that in the $d_{x^2-y^2}$+$s$-wave case, 
$\langle \Delta_{\bf k}\rangle\neq 0$ and
$\Sigma_1$ in Eq.~(\ref{eq:Sigmas.1}) is nonzero
and must be retained in the calculation.
These complications are accounted for (see Ref.~\cite{Wu97})
in obtaining the above result.
Moreover, in this case, the second (screening) term of
Eq.~(\ref{eq:chi.Raman.sc.simplified}) contributes in 
the $B_{1g}$ channel (but not in the $B_{2g}$ channel),
though this contribution is expected to be small
when the $s$-wave component is small.

In conclusion, we have found that channel-dependent Raman scattering
can be used to test a universal low-temperature behaviour in
disorder high-$T_c$ superconductors which in turn can reveal
information on the pairing state in these materials. We studied the slopes
of low-temperature Raman intensity at zero frequency in various channel.
Similar to what was found
in the microwave or dc conductivity \cite{Lee93} and thermal 
conductivity \cite{GYSR96},  for a $d_{x^2-y^2}$-wave 
superconductor with nodal order parameter along the diagonals
on the Fermi surface, the universal feature
holds in both $A_{1g}$- and $B_{2g}$-channel Raman slopes.
Such a feature is not found, however,  in the $B_{1g}$-channel Raman slope.
Moreover, in contrast to microwave or thermal conductivity where
the universal features are general to all the pairing states 
with nodes and certain reflection symmetry,
the channel-dependent universality or lack thereof in Raman scattering 
is sensitive to difference in pairing states and hence allows one
to further clarify the symmetry of the pairing states. 
In addition for a mixed order parameter with $d_{x^2-y^2}$
and $s$ symmetries, we have shown that Raman is able to
give the amount of $s$-wave component which is not the case
using microwave or thermal conductivity.
Finally, we remark that, as recently shown by Branch \cite{Branch},
the strong inelastic scattering effect in the high-$T_c$ materials
may dominate over the elastic impurity scattering but will not
destroy the universality in Raman slopes predicted here
because this is a zero-temperature effect and at $T=0$ the inelastic 
scattering time becomes infinite. 

We are grateful to D. Branch for enlightening discussions.
This work was supported by Natural Sciences and Engineering Research Council
(NSERC) of Canada and Canadian Institute for Advanced Research (CIAR).
 

\end{document}